\documentclass[11pt, a4paper]{article}
 \evensidemargin \oddsidemargin

\usepackage{graphicx}
\usepackage{setspace}
\usepackage[version=3]{mhchem} %
\usepackage[T1]{fontenc}       %
\usepackage{amsmath}
\usepackage{amssymb}
\usepackage{bm}
\usepackage{cite}
\usepackage{url,here}
\usepackage[version=3]{mhchem}
\author{Yutaka Hori$^*$, Hiroki Miyazako$^*$}
\title{Analyzing Diffusion and Flow-driven Instability\\ using Semidefinite Programming}%

\date{}
\begin{document}
\maketitle
{\let\thefootnote\relax\footnote{{
* The authors contributed equally to this work.\\
Yutaka Hori is with Department of Applied Physics and
Physico-Informatics, Keio University. 3-14-1 Hiyoshi, Kohoku-ku,
Yokohama, Kanagawa 223-8522, Japan.\\
Hiroki Miyazako is with Department of Information Physics and Computing,
The University of Tokyo. 7-3-1 Hongo, Bunkyo-ku, Tokyo 113-8656, Japan.\\
Correspondence should be addressed to Yutaka Hori (yhori@appi.keio.ac.jp).
This work should be cited as Y. Hori, H. Miyazako, ``Analysing diffusion and flow-driven instability using semidefinite programming,'' {\it Journal of the Royal Society Interface}, Vol. 16, No. 150, 20180586, 2019. \url{https://doi.org/10.1098/rsif.2018.0586}
}}}

 \begin{abstract}
Diffusion and flow-driven instability, or transport-driven instability, is one of the central mechanisms
to generate inhomogeneous gradient of concentrations in spatially distributed chemical systems. %
However, verifying the transport-driven instability of reaction-diffusion-advection systems 
 requires checking the Jacobian eigenvalues of infinitely many Fourier modes, which is computationally intractable.
To overcome this limitation, this paper proposes mathematical
  optimization algorithms that determine the stability/instability of
  reaction-diffusion-advection systems by finite steps of algebraic
  calculations.
Specifically, the stability/instability analysis of Fourier modes is
  formulated as a sum-of-squares (SOS) optimization program, which is a
  class of convex optimization whose solvers are widely available as
  software packages.
 The optimization program is further extended for facile
  computation of the destabilizing spatial modes. This extension allows for predicting and designing the shape of concentration gradient without
  simulating the  governing equations.
The streamlined analysis process of self-organized pattern
  formation is demonstrated with a simple illustrative reaction model with diffusion and advection.

 \end{abstract}
\clearpage

\section{Introduction}
Molecular transportation is a fundamental mechanism that couples spatially 
distributed chemical reaction networks and enables self-organized 
pattern formation in biology and chemistry. 
For example, in developmental biology, passive diffusion and advective
flow of signaling molecules within and between cells are known to play a central role
in identifying and regulating positions and shapes during %
embryo-genesis \cite{Koch1994,Murray2003, Kondo2010} and cell division
\cite{Hu1999,Meinhardt2001,Lutkenhaus2007,Rudner2010,Vecchiarelli2016}. %
In chemistry, diffusion-driven self-organized patterns were 
reconstituted in crafted reactors with commodity chemicals \cite{Castets1990,Ouyang1991}. 
More recently, synthetic biologists have been attempting to utilize
quorum sensing, a mechanism of diffusion-based cell-to-cell signaling of
E. coli, to regulate a population of synthetic biomolecular reaction systems that communicate \cite{You2004, Basu2005,Liu2011,Moon2012,Tamsir2011} and synchronize with each other \cite{Danino2010,Din2016,Scott2017,Baumgart2017}.

\par
\smallskip
The theoretical foundation of transport-driven self-organization was established by Turing \cite{Turing1952}, where he showed using a reaction-diffusion equation that molecular
concentrations can form a spatially periodic gradient due to the interaction of reaction and diffusion 
despite the averaging nature of diffusion.  
Later, Rovinsky and Menzinger \cite{Rovinsky1992,Rovinsky1993} verified
both mathematically and experimentally that differential flow, or advection, of molecules can also induce an oscillatory spatio-temporal
concentration gradient. 
In these works, mathematical analyses revealed that the pattern
 formation was due to the destabilization of spatial oscillation modes,
 and the destabilization was induced by the difference of diffusion and flow rates between
 reactive molecules. Mathematically, this can be interpreted that molecular transportation caused
 spontaneous growth of some spatial Fourier modes and led to
 spatially periodic pattern formation at steady state. %

\par
\smallskip
Currently, a widely used approach to analyzing the transport-driven
pattern formation is the linear stability/instability analysis of spatial Fourier
components. %
Specifically, the Jacobian eigenvalues of the governing reaction-diffusion-advection
equations are computed for different spatial Fourier modes to explore the
existence of unstable harmonic components.
Then, spatially periodic pattern formation is expected if an
eigenvalue of the Jacobian resides in the right-half complex plane for
some non-zero frequency components.
For relatively simple reaction systems, analytic instability conditions
were obtained using this approach, and the parameter space for pattern formation was
thoroughly characterized  \cite{Gierer1972, Koch1994, Hori2015}. 
For large-scale and complex reaction systems, on the other hand, instability analysis 
requires substantial computational efforts to iteratively compute Jacobian eigenvalues for each Fourier component.
However, this approach is essentially incapable of drawing mathematically rigorous conclusion on the stability of 
reaction-diffusion-advection systems since the spatial gradient of chemical concentrations is represented with
infinitely many Fourier modes.

\par
\smallskip
Another approach to analyzing the stability/instability of reaction-diffusion equations is to 
use Lyapunov's method, where the core idea is to guarantee the energy dissipation of certain Fourier
components by constructing a Lyapunov function. %
The exploration of a Lyapunov function typically reduces to an 
algebraic optimization problem called semidefinite programming (SDP), 
which is an efficiently solvable class of convex optimization \cite{Boyd1994,Boyd2008}.
Existing works presented sufficient conditions for the stability of reaction-diffusion equations, certifying 
non-existence of spatially inhomogeneous solutions \cite{Jovanovic2008,Arcak2011,Shafi2013}. 
A more general SDP based framework was also developed for the stability analysis of univariate nonlinear partial differential equations by exploring polynomial integral Lyapunov functions \cite{Valmorbida2016}. 
However, the focus of these works is spatially homogeneous behavior, 
and the conditions are not necessarily suitable for studying diffusion-driven instability that leads to spatially periodic pattern formation.

\par
\smallskip
This paper presents computationally tractable necessary and sufficient
conditions for the local stability/instability of reaction-diffusion-advection
systems. The proposed approach is amenable to computational
implementation and is particularly convenient for analyzing
transport-driven instability. 
Specifically, we derive linear matrix inequality (LMI) conditions that certify the stability/instability of infinitely many Fourier
 components. This algebraic condition can be efficiently verified with
 semidefinite programming (SDP) \cite{Boyd1994,Boyd2008}. %
The derivation of these conditions is based on the linear stability analysis of Fourier
components. 
Although the analysis essentially requires checking the roots of infinitely many characteristic polynomials with complex
coefficients, we show that the %
local stability is equivalent to an existence of sum-of-squares
(SOS) decomposition of certain Hurwitz polynomials, which can be
formulated as a semidefinite optimization problem \cite{Parrilo2003}.
Based on this result, we further derive conditions for certifying the stability/instability of reaction-diffusion-advection systems for a prespecified
 set of spatial frequency. %
This extension allows for facile computation of the destabilizing
spatial modes, enabling the prediction of spatial oscillations without
simulating the governing equation.
The proposed instability analysis helps not only better
understanding of reaction-diffusion-advection kinetics but also 
engineering of chemical reactions in synthetic biology and chemistry.

\par
\smallskip
The following notations are used in this paper. 
$\mathbb{N}$ is a set of positive integers.
$\mathbb{Z}$ is a set of all integers.
$\mathbb{R}$ is a set of real numbers. $\mathbb{R}^{{n} \times {n}}$ is a
set of $n$ by $n$ matrices with real entries. $|A|$ is a determinant of
the matrix $A$.
$A \succeq O$ means that $A$ is positive semidefinite.
$\mathrm{deg}(p(x))$ is the degree of a polynomial $p(x)$. $\lceil x
\rceil$ is the ceiling function, that is, the smallest integer that is
greater than or equal to $x$.

\section{Model of reaction-diffusion-advection system}
We consider a reaction-diffusion-advection process of $n$
molecular species %
 defined in a finite spatial domain $\Omega$.
For notational simplicity, we consider only one dimensional space $\Omega := [0,L]$, whose coordinate
is specified by the symbol $x$, but all
theoretical results shown in this paper can be generalized to higher dimensions.
Let $C_i(x, t)$ denote the concentration of %
 the $i$-th molecule at  position $x \in \Omega$ and time $t$ and ${\bm C}(x, t)$ be 
the vector of the molecular concentrations
${\bm C}(x, t) := [C_1(x,t), C_2(x,t), \cdots, C_{n}(x,t)]^T$. 
The spatio-temporal dynamics of the molecular concentrations are then described by 
the reaction-diffusion-advection equation
 \begin{align}
&\frac{\partial {\bm C}(x, t)}{\partial t} = 
	{\bm f}({\bm C}(x, t))
	+  D \frac{\partial^2 {\bm C}(x,t)}{\partial x^2} + V
   \frac{\partial {\bm C}(x,t)}{\partial x},
\label{nonlin-rdae}
 \end{align}
 where the function ${\bm f}(\cdot)$ is a $C^{1}$ vector-valued function governing
 local reactions, and $D := \mathrm{diag}(d_1, d_2, \cdots, d_n)$ and $V :=
 \mathrm{diag}(v_1, v_2, \cdots, v_n)$ are the coefficients of diffusion and flow
 velocity, respectively. We assume $d_i > 0$ and $v_i \ge 0$ in the
 following theoretical development $(i=1,2,\cdots,n)$.

\par
\begin{figure}[tb]
  \centering
\includegraphics[clip,width=14cm]{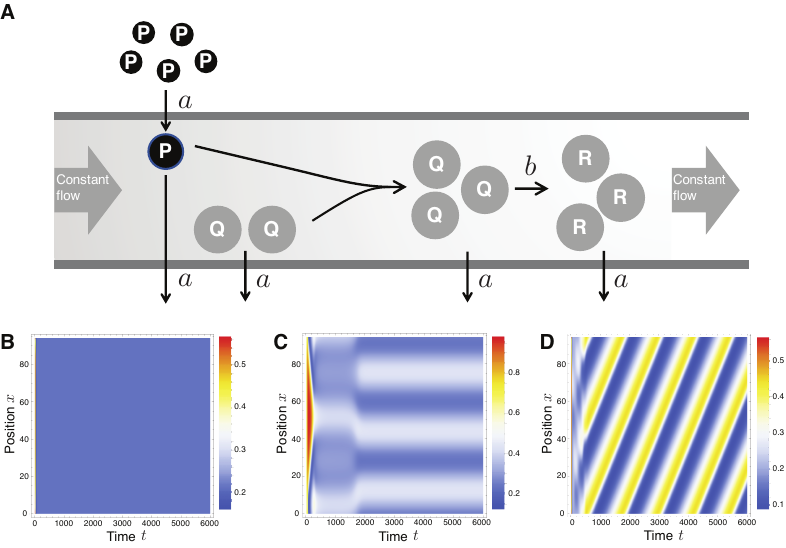}
  \caption{{\bf A reaction-diffusion-advection system inducing
  flow-driven instability} (A) Schematics of the reaction model with
  diffusion and advection. (B-D) Spatio-temporal dynamics of concentration
  gradient for different values of parameters.}
  \label{sim-fig}
 \end{figure}

\smallskip
The reaction-diffusion-advection system shows a variety of
spatio-temporal dynamics ranging from spatially uniform steady state to
spatio-temporal oscillations depending on the parameters and the stoichiometry of the reactions.
As a motivating example, %
we consider the following set of reactions that
consists of molecules $P$, $Q$, and a product $R$: 
\begin{align}
   \ce{P + 2Q -> 3Q},
~~~~   \ce{Q -> R}, 
   \label{reaction-eq}
  \end{align}
where the molecule $Q$ catalyzes its own production using the substrate $P$, and the product $R$ is inert to
the reactions \cite{Gray1984,Pearson1993}.
The substrate $P$ is constantly supplied at a constant rate and all molecules
 are drained at the same rate as illustrated in Fig. \ref{sim-fig}(A).
We assume that all molecules are spatially distributed in one
dimensional space $\Omega := [0,30\pi]$ with periodic boundary conditions, and they are transported by constant flow and passive diffusion.
The spatio-temporal dynamics of the molecular concentrations are then modeled by 
\begin{equation}
\begin{array}{l}
\displaystyle  \frac{\partial C_1}{\partial t} =  -C_1 C_2^{2}+a(1-C_1)+d
  \frac{\partial^2 C_1}{\partial x^2} + 
  v_1 \frac{\partial C_1}{\partial x},\\
\displaystyle \frac{\partial C_2}{\partial t}  =  C_1
  C_2^{2}-(a+b)C_2+\frac{\partial^2 C_2}{\partial x^2} + 
  v_2 \frac{\partial C_2}{\partial x},
\end{array}
  \label{gray-scott-model}
\end{equation}
 where $C_1(x, t)$ and $C_2(x, t)$ denote the concentrations of $P$ and $Q$, respectively.
 In the model (\ref{gray-scott-model}), the reaction rates are
 normalized by that of the autocatalytic reaction in (\ref{reaction-eq}). The constant $a$ and $b$ represent the normalized
 supply rate of the substrate $P$ and the production rate of $R$ (see
 Fig. \ref{sim-fig}(A)). The spatial coordinate $x$ is defined so that
 the diffusion coefficient of the molecule $Q$ becomes one.

\par
\smallskip
Fig. \ref{sim-fig}(B)-(D) illustrate qualitatively different
spatio-temporal dynamics for the reaction-diffusion-advection system
(\ref{gray-scott-model}) for different choices of $b$, $v_1$ and $v_2$. 
Specifically, $(b, v_1, v_2) = (0.040, 0, 0)$ for Fig. \ref{sim-fig}(B),
$(b, v_1, v_2) = (0.055, 0, 0)$ for Fig. \ref{sim-fig}(C), and
$(b, v_1, v_2) = (0.040, 1.897, 0.3162)$ for Fig. \ref{sim-fig}(D) were used
(see Method section for the other parameters and initial values). 
The concentration of $P$ forms spatially periodic
oscillations as $b$ increases from $0.040$ to $0.055$ 
despite the averaging effect of the passive diffusion (Fig. \ref{sim-fig}(B), (C)).
This pattern formation, which is widely known as Turing pattern formation, is caused by the
destabilization of certain spatial oscillation modes due to the difference
of the diffusion rates (diffusion instability) \cite{Murray2003}.
On the other hand, %
the spatio-temporal oscillations in Fig. \ref{sim-fig}(D) is induced by 
a different destabilization mechanism based on the advective transportation of the molecules.

\par
\smallskip
In the following sections, we first review that these destabilizing 
effects can be explained by probing the local instability of Fourier modes of the reaction-diffusion-advection system. 
We then present novel algebraic stability conditions for the stability/instability analysis of infinitely many Fourier modes with semidefinite programming.

\section{Stability analysis of spatial Fourier components}
\par
\smallskip
To analyze instability of spatial modes associated with spatial pattern
formation, we linearize the equation (\ref{nonlin-rdae}) around a spatially homogeneous equilibrium
point ${\bm C}(x,t) = \bar{{\bm C}}$. 
In other words, $\bar{{\bm C}}$ is an equilibrium of local reactions
satisfying  ${\bm f}(\bar{{\bm C}}) = 0$. %
Assuming the existence of such equilibrium point,
 we can write the evolution of the molecular concentrations around $\bar{{\bm C}}$ by 
\begin{align}
\frac{\partial {\bm c}(x, t)}{\partial t} = 
A {\bm c}(x,t)
 + 
D
\frac{\partial^2{\bm c}(x,t)}{\partial x^2}
+
V
\frac{\partial{\bm c}(x,t)}{\partial x},
\label{linear-eq}
\end{align}
where ${\bm c}(x, t) := {\bm C}(x, t) - \bar{\bm C}$ is the vector of relative 
concentrations, and $A$ is the Jacobian of ${\bm f}(\cdot)$ evaluated at
$\bar{{\bm C}}$.

\par
\smallskip
The local behavior of ${\bm c}(x, t)$ can be characterized by its spatial Fourier components $\tilde{\bm c}(\zeta, t)$ 
when the boundaries of the spatial domain $\partial \Omega$ satisfy certain conditions.
Specifically, let $\tilde{\bm c}(\zeta,t)$ be the Fourier coefficients of ${\bm c}(x,t)$ satisfying %
\begin{align}
{\bm c}(x, t) = \sum_{\zeta \in \mathcal{Z}} \tilde{\bm c}(\zeta, t) e^{j \zeta x},
\label{tildec-def}
\end{align}
where $\mathcal{Z}$ is a set of discrete frequency variables that depends on the boundary conditions (see Remark 1).
Multiplying by $e^{-j\zeta x}$ and taking the integral of both sides of
(\ref{linear-eq}), we have 
\begin{align}
\frac{d \tilde{{\bm c}}(\zeta, t)}{d t} =  (A - 
\zeta^2D + j\zeta {V}) \tilde{{\bm c}}(\zeta, t). 
\label{freq-eq}
\end{align}
This equation represents the dynamics of frequency component
$\tilde{{\bm c}}(\zeta, t)$.
It should be noticed that, for each fixed $\zeta$, (\ref{freq-eq}) is an $n$-th order linear time-invariant
ordinary differential equation (ODE) with complex coefficients. 
Thus, the reaction-diffusion-advection equation is decomposed
into a set of infinitely many ODEs. 

\medskip
\noindent
{\bf Remark 1.}
The frequency variable $\zeta$ in (\ref{tildec-def}) takes discrete values  that depend on the boundary conditions. 
Specifically, let the set of all frequency variables for a given boundary condition be denoted by $\mathcal{Z}$. 
Then, $\mathcal{Z} = \{2k \pi /L \}_{k \in \mathbb{Z}}$ for periodic boundary conditions. 
When $V = O$, $\mathcal{Z}$ is defined by $\mathcal{Z} = \{k\pi /L\}_{k \in \mathbb{Z}}$ for the Neumann boundary condition and $\mathcal{Z} = \{k \pi /L\}_{k \in \mathbb{Z} \backslash \{0\}}$ for the Dirichlet boundary condition, respectively \cite{Strauss2007}. 
The reader is referred to \cite{Strauss2007} for other boundary conditions. 
It should also be noted that $\mathrm{Im}[\tilde{\bm c}(\zeta, t)] = 0$ for the Neumann boundary with $V = 0$ and 
$\mathrm{Re}[\tilde{\bm c}(\zeta, t)] = 0$ for the Dirichlet boundary with $V = 0$, implying that the Fourier cosine and sine transforms 
are used to obtain $\tilde{\bm c}(\zeta, t)$, respectively.

\par
\medskip
Since the equation (\ref{freq-eq}) represents the dynamics of Fourier
components, ${\bm c}(x, t)$ asymptotically converges to zero
if the growth rate of  $\tilde{{\bm c}}(\zeta, t)$
 is negative for all frequency components $\zeta$. 
On the other hand, if there exists a frequency component
$\tilde{{\bm c}}(\zeta, t)$ with non-zero frequency $\zeta$ whose growth rate is positive, the corresponding non-zero spatial mode is 
 amplified around the spatially homogeneous equilibrium $\bar{{\bm C}}$. 
In the nonlinear reaction-diffusion-advection system (\ref{nonlin-rdae}), the unstable frequency component potentially induces periodic pattern formation in space. 
In other words, the existence of a growing frequency component with non-zero frequency is a necessary condition for the generation of spatially periodic pattern in the system (\ref{nonlin-rdae}).

\par
\smallskip
More formally, this can be stated as local stability/instability of the nonlinear reaction-diffusion-advection system (\ref{nonlin-rdae}).
The nonlinear system (\ref{nonlin-rdae}) is said to be locally stable around $\bar{{\bm C}}$ if the linear system (\ref{freq-eq}) is asymptotically stable.
For linear systems, asymptotic stability is determined from the eigenvalues of the matrix $A-\zeta^2 D + j\zeta V$. 
More specifically, the following lemma holds (see Supplementary Information for the proof).

\medskip
\noindent
{\bf Lemma 1.}
Consider the linear reaction-diffusion-advection system (\ref{linear-eq}).
The system (\ref{linear-eq}) is 
asymptotically stable
if and only if
 all roots of the characteristic polynomial
\begin{align}
 \varphi(\zeta,s) := |sI - A + D\zeta^2 - j\zeta V | = 0
 \label{varphi-def}
\end{align}
are in the open left-half complex plane $\{s \in \mathbb{C}~|~\mathrm{Re}[s] < 0\}$ for all 
$\zeta \in \mathcal{Z}$, where %
$\mathcal{Z}$ is defined in Remark 1.

\par
\medskip
Thus, the stability analysis of the reaction-diffusion-advection system (\ref{linear-eq}) reduces to 
finding the roots of the polynomial $\varphi(\zeta, s) = 0$ for $\zeta \in \mathcal{Z}$.
It should, however, be noted that verifying the stability condition requires solving infinitely many polynomial equations with complex
coefficients (\ref{varphi-def}), and analytic conditions are hardly obtained except for some special cases \cite{Rovinsky1992,Bamforth2001}. 
Consequently, most of the existing works resort to approximate stability analysis by %
solving (\ref{varphi-def}) for a finite range of quantized $\zeta$.
In the next section, we propose a novel computational approach that
overcomes this limitation. Specifically, we introduce a computationally tractable optimization
problem that certifies the stability of the reaction-diffusion-advection
system. %

\section{Linear matrix inequality condition for stability/instability analysis}
  \label{LMI-sec}
In this section, we show matrix inequality conditions for computational verification of the stability/instability of the system (\ref{linear-eq}).
For theoretical development, we here consider a slightly modified condition of Lemma 1 that 
all roots of the polynomial (\ref{varphi-def}) lie in the open left-half complex plane for $\zeta \in \mathbb{R}$ instead of $\zeta \in \mathcal{Z}$. 
Mathematically, this becomes only a sufficient condition for the asymptotic stability of the system (\ref{linear-eq}). 
However, conventionally, this condition has been used for the stability/instability analysis of reaction-diffusion-advection systems \cite{Turing1952,Rovinsky1992,Rovinsky1993} since 
the elements of $\mathcal{Z}$, which lie on the real axis $(-\infty, \infty)$ tend to be close to each other as the length of the domain $L$  becomes large, and the gap from the necessary condition becomes small enough in practical applications.

\par
\smallskip
To this end, we first review an algebraic stability condition due to
Hurwitz (Theorem 3.4.68 of \cite{Hinrichsen2011}).
Let $\varphi_{\rm Re}(\zeta, s)$ and $\varphi_{\rm Im}(\zeta, s)$ be
real and imaginary parts of the complex polynomial $\varphi(\zeta, js)$, respectively.
Specifically, we denote by 
\begin{align}
\varphi(\zeta,js) = \varphi_{\rm Re}(\zeta, s) + j \varphi_{\rm
 Im}(\zeta, s)
 \label{varphi-def2}
\end{align}
with
\begin{align}
&\varphi_{\rm Re}(\zeta,s) = p_{n}(\zeta)s^n + p_{n-1}(\zeta) s^{n-1} +
 p_{n-2}(\zeta) s^{n-2} + \cdots + p_{0}(\zeta),\notag \\
&\varphi_{\rm Im}(\zeta,s) = q_{n}(\zeta)s^{n} + q_{n-1}(\zeta) s^{n-1} +
 q_{n-2}(\zeta) s^{n-2} + \cdots + q_{0}(\zeta),\notag
\end{align}
where $p_i(\zeta)$ and $q_i(\zeta)$ are polynomials of $\zeta$ with real
coefficients.
Note that $\varphi_{{\rm Re}}(\zeta, s)$ and $\varphi_{{\rm Im}}(\zeta,
s)$ are not defined for $\varphi(\zeta, s)$ but for $\varphi(\zeta,
js)$ in (\ref{varphi-def2}).
Using these polynomials, we define the following $2n \times 2n$ Sylvester matrix 
  \begin{align}
S:=\begin{bmatrix}
q_{n} & q_{n-1} & q_{n-2} & \cdots & \cdots & q_{0} & 0 & \cdots  & 0\\
p_{n} & p_{n-1} & p_{n-2} & \cdots & \cdots & p_{0} & 0 & \cdots & 0\\
0 & q_{n} & q_{n-1} & q_{n-2} & \cdots & \cdots &  q_{0} & \cdots 
 & 0\\
0 & p_{n} & p_{n-1} & p_{n-2} & \cdots & \cdots & p_{0} & \cdots 
 & 0\\
\vdots & \ddots & \ddots & \ddots & \ddots & \ddots &\ddots  & \ddots
 & \vdots\\
0 & \cdots & \cdots & \cdots & q_{n} & q_{n-1} & q_{n-2} & \cdots 
 & q_0\\
 0 & \cdots & \cdots & \cdots & p_{n} & p_{n-1} & p_{n-2} & \cdots 
 & p_0
   \end{bmatrix}
   \label{S-def}
  \end{align}
and define $2i$-th order principal minors of
the matrix by $\Delta_i(\zeta)~(i=1,2,\cdots,n)$. For example, 
\begin{align}
 \Delta_1(\zeta) =
 \left|
\begin{array}{cccc}
 q_{n} & q_{n-1} \\
 p_{n} & p_{n-1}
\end{array}
\right|
 ,~
 \Delta_2(\zeta) =
 \left|
\begin{array}{cccc}
 q_{n} & q_{n-1} & q_{n-2} & q_{n-3} \\
 p_{n} & p_{n-1} & p_{n-2} & p_{n-3} \\
  0 & q_{n} & q_{n-1} & q_{n-2} \\
 0 &  p_{n} & p_{n-1} & p_{n-2}
\end{array}
 \right|
 \notag
\end{align}
and $\Delta_n = |S|$.
According to the Hurwitz criterion for complex polynomials 
(Theorem 3.4.68 of \cite{Hinrichsen2011}), all roots of $\varphi(\zeta, s) = 0$ lie in the
open left-half complex plane for a given $\zeta$ if and only if the
coefficients satisfy the polynomial inequalities $\Delta_i(\zeta) >
0~(i=1,2,\cdots,n)$. 
Thus, for each fixed value of $\zeta$, we can check the stability of the
corresponding Fourier mode using computationally tractable algebraic conditions.
To analyze the stability of the reaction-diffusion-advection system
(\ref{linear-eq}), however, we need to guarantee the stability for all
frequency $\zeta$, which is computationally intractable.
As a result, the stability analysis of the reaction-diffusion-advection
system often resorts to approximation by iteratively checking the sign
of $\Delta_i(\zeta)$ for a finite range of discretized values of $\zeta$.
To overcome this issue, we introduce a computationally tractable 
condition for certifying non-negativity of $\Delta_i(\zeta)$.

\medskip
\noindent
{\bf Proposition 1.}
Let $M_i \in \mathbb{R}^{(\ell_i+1) \times (\ell_i+1)}$ be any one of the matrices satisfying
\begin{align}
\Delta_i(\zeta) = {\bm z}_i^T M_i {\bm z}_i,
\label{delta-def}
\end{align}
where $\ell_i := \lceil{\mathrm{deg}(\Delta_i(\zeta))/2}\rceil$ and ${\bm z}_i := [1, \zeta, \zeta^2, \cdots, \zeta^{\ell_i}]^T \in
\mathbb{R}^{\ell_i + 1}~(i=1,2,\cdots,n)$. Then, the following (i) and (ii) are equivalent.
 \begin{enumerate}
\item[(i)] $\Delta_i(\zeta) \ge 0$ for all $\zeta \in \mathbb{R}$ and $i=1,2,\cdots,n$.
 \item[(ii)] There exists a symmetric matrix $N_i \in \mathbb{R}^{(\ell_i+1) \times (\ell_i+1)}$ such that
	      \begin{align}
	      M_i + N_i \succeq 0~~{\rm and}~~ \sum_{(j, k) \in \Theta_{\ell}}  \nu_{jk}^{(i)} = 0 \label{MN-cond}
	      \end{align}
	      for $\ell=2,3,\cdots,2 \ell_i+2$ and $i=1,2,\cdots,n$, 
	      where $\nu_{jk}^{(i)}$ is the $(j, k)$-th entry of the matrix
	      $N_i$, and
\begin{align}
\Theta_\ell := \{(j, k) \in \mathbb{N} \times
	      \mathbb{N}~|~ j + k = \ell, 1 \le j \le \lceil \ell/2 \rceil, 1 \le k
	      \le \lceil \ell/2 \rceil \}.
\end{align}
\end{enumerate}

\par
\medskip
The core idea of this proposition is to find a sum of squares (SOS)
decomposition of $\Delta_i(\zeta)$ (see Supplementary Information for
the proof).
That is, we certify non-negativity of $\Delta_i(\zeta)$ by showing that
$\Delta_i(\zeta)$ can be represented as a sum of non-negative terms.
In general, we can always find a constant matrix $M_i$ that makes the quadratic form (\ref{delta-def}) for a 
given polynomial $\Delta_i(\zeta)$. 
Thus, a sufficient condition for the non-negativity of the polynomial $\Delta_i(\zeta)$ is $M_i \succeq O$. 
However, this does not constitute a necessary condition since the choice of $M_i$ is not unique. 
In other words, we need to explore all possible quadratic expressions of $\Delta_i(\zeta)$ to show the necessary condition.
The matrix $N_i$ satisfying ${\bm z}_i^T N_i {\bm z}_i = 0$ is added to $M_i$ in the condition (ii) for this purpose.

\par
\smallskip
The condition (ii) is amenable to computational implementation since
 it requires finding a single set of matrices $N_i~(i=1,2,\cdots,n)$
 that satisfies (\ref{MN-cond}) instead of verifying the non-negativity
 of $\Delta_i(\zeta)$ for all $\zeta$. %
In fact, the problem of finding $N_i$ in (\ref{MN-cond}) can be reduced to semidefinite programming (SDP) \cite{Boyd1994,Boyd2008},  which is a class of convex optimization program with a linear objective function and
 semidefinite constraints. 
Thus, we can utilize existing SDP solvers such as SeDuMi \cite{Sturm1999} and
 SDPT3 \cite{Toh1999}, which implement interior point methods \cite{Forsgren2002} to efficiently search for the matrix $N_i$. 
It should be noted that the linear equality constraints in (\ref{MN-cond}) can be equivalently transformed to semidefinite
constraints $L \succeq O$ and $L \preceq O$ with a diagonal matrix
 $L$ whose entries are the left-hand side of the equality constraints,
 {\it i.e.,} $L:= \mathrm{diag}(\sum_{(j,k)\in\Theta_2} \nu_{jk}^{(1)},
 \sum_{(j,k)\in\Theta_3} \nu_{jk}^{(1)}, \cdots,
 \sum_{(j,k)\in\Theta_{2\ell_{n}}} \nu_{jk}^{(n)})$. 
 Thus, all of the constraints in the condition (ii) can be reduced to semidefinite conditions.

 \medskip
 \noindent

\medskip
\noindent
{\bf Remark 2.}~
Proposition 1 can be also used for the stability analysis of reaction-diffusion-advection process in multi-dimensional space $\Omega$. 
To see this, we consider $m$ dimensional space $\Omega$, and define a vector of spatial frequency ${\bm \zeta} := [\zeta_1, \zeta_2, \cdots, \zeta_m]^T$. 
Then, the dynamics of multi-dimensional Fourier components $\tilde{\bm c}({\bm \zeta}, t)$ is obtained as 
\begin{align}
\frac{d \tilde{{\bm c}}({\bm \zeta}, t)}{d t} =  (A - \bar{\zeta}^2D + j\bar{\zeta} {V}) \tilde{{\bm c}}({\bm \zeta}, t), 
\end{align}
where $\bar{\zeta} := \sum_{i=1}^{m} {\zeta}_i$. 
This leads to the characteristic polynomial that determines the stability of the reaction-diffusion-advection system as 
$\varphi({\bm \zeta}, s) = |sI - A  + \bar{\zeta}^2 D - j \bar{\zeta} V|$. 
Here, an important implication is that $\{A - \bar{\zeta}^2D + j\bar{\zeta} {V}~|~\bar{\zeta} := \sum_{i=1}^{m} \zeta_i,~~\zeta_i \in \mathbb{R}\} = \{A - \zeta^2 D + j \zeta {V}~|~\zeta \in \mathbb{R}\}$.
Thus, $\varphi({\bm \zeta}, s) = 0$ has all roots in the open left-half complex plane for all ${\bm \zeta} \in \mathbb{R}^m$ if and only if 
$\varphi(\zeta, s) = 0$ does so for all $\zeta \in \mathbb{R}$. 
The latter condition boils down to checking the non-negativity of $\Delta_i(\zeta)$ as discussed in this section.
Therefore, the matrix inequality conditions in Proposition 1 can be used for the stability analysis of multi-dimensional reaction-diffusion-advection systems.

  \begin{figure}[tb]
  \centering
\includegraphics[clip,width=14cm]{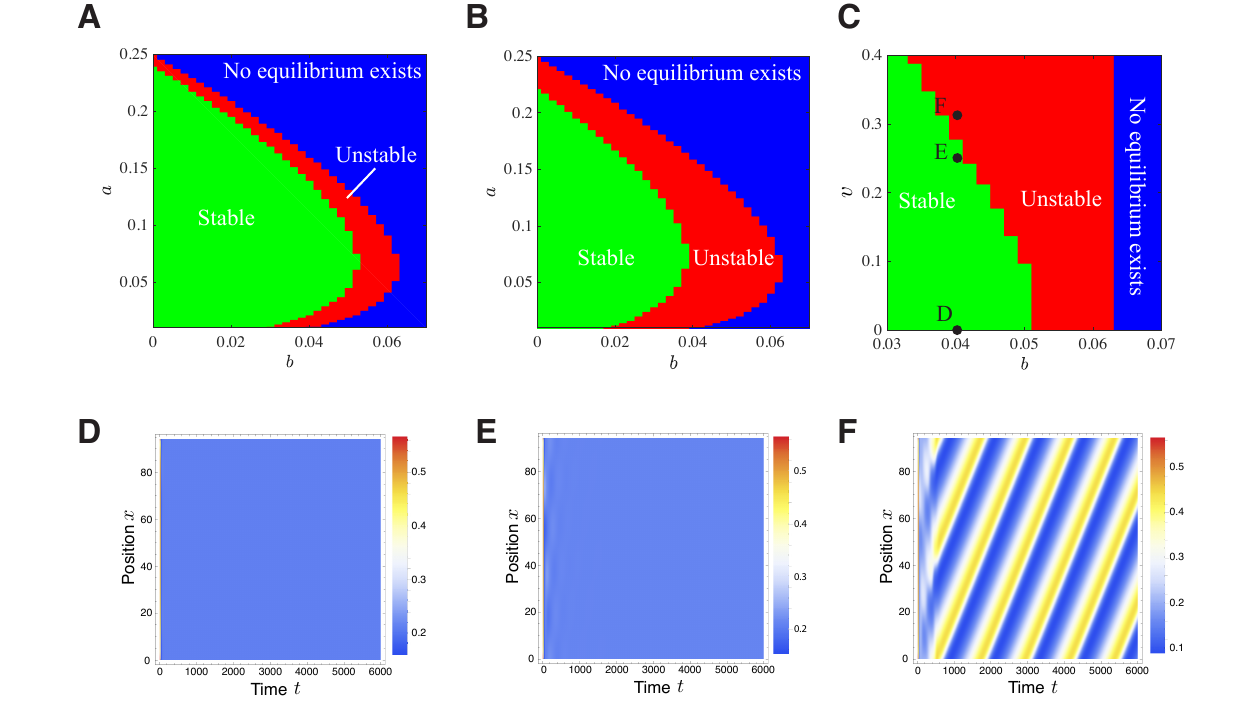}
  \caption{{\bf Parameter space analysis for reaction-diffusion-advection system
  (\ref{gray-scott-model})}.
  (A, B) Parameter maps showing stable and
  unstable regions without advection ($v = 0$) and with advection ($v =
  0.3162$). (C) Parameter map showing the robustification of transport-driven
  instability. (D-F) Spatio-temporal profiles of the molecule $P$ for
  different values of parameters (see Fig. 2 (C)). The panels D and F
  are repeated from Fig. 1 (B) and Fig. 1 (D), respectively.}
  \label{pmap}
 \end{figure}

\medskip
\noindent
{\bf Example. }
We demonstrate the SDP based stability test using the reaction-diffusion-advection system (\ref{gray-scott-model}). 
Let $[C_1^*, C_2^*]^T$ be a spatially homogeneous equilibrium point of the system and consider the linearized system
 (\ref{linear-eq}).
It follows from (\ref{gray-scott-model}) that 
\begin{align}
 C_{1}^*=\frac{1}{2}(1-\sqrt{w}),\, C_{2}^*=\frac{a}{2(a+b)}(1+\sqrt{w})
\end{align}
is an equilibrium point, where $w:=1-4(a+b)^{2}/a$.
The Jacobian linearization leads to 
  \begin{align}
   \frac{\partial}{\partial t}
\begin{bmatrix}
  c_1 \\
 c_2
\end{bmatrix}
 = 
\begin{bmatrix}
 -a-C_{2}^{*2} & -2C_{1}^*C_{2}^*\\
 C_{2}^{*2} & -(a+b)+2C_{1}^*C_{2}^*
\end{bmatrix}
\begin{bmatrix}
 c_1  \\
 c_2
\end{bmatrix}
+ 
 \begin{bmatrix}
d & 0\\
 0 & 1
 \end{bmatrix}
\begin{bmatrix}
\nabla^2 c_1\\
\nabla^2 c_2 
\end{bmatrix}
 + 
 \begin{bmatrix}
  v_1  & 0\\
   0 & v_2
 \end{bmatrix}
\begin{bmatrix}
 \nabla c_1  \\
 \nabla c_2
\end{bmatrix}
   .
\label{linear-rd-eq}
\end{align}

\par
\smallskip
To analyze the stability of the equilibrium point, %
we substitute the parameters into (\ref{varphi-def})
and compute the polynomials $\varphi_{\rm Re}(\zeta, s)$ and
$\varphi_{\rm Im}(\zeta, s)$, and define  the Sylvester matrix $S$
defined in (\ref{S-def}).

\par
\smallskip
Here we first consider the parameter values used in Fig. \ref{sim-fig}(B), where $(b, v_1, v_2) = (0.040, 0, 0)$ (see Method section for the other parameters).
The corresponding polynomials $\Delta_i(\zeta)~(i=1,2)$ are then obtained as 
\begin{align}
& \Delta_1(\zeta) = 7\zeta^2 + 0.184, \label{delta1-eq} \\
& \Delta_2(\zeta) = 294\zeta^8-0.0382 \zeta^6 + 0.192 
 \zeta^4+0.0315 \zeta^2 + 0.000555. \label{delta2-eq}
\end{align}
Hurwitz's stability condition implies that the equilibrium $[C_1^*,
C_2^*]$ is stable if and only if $\Delta_1(\zeta) > 0$ and
$\Delta_2(\zeta) > 0$ for all $\zeta \in \mathbb{R}$.
It is obvious that $\Delta_1(\zeta) > 0$, but the sign of 
$\Delta_2(\zeta)$ needs further examination.
Thus, we use the condition (ii) of Proposition 1 to derive conditions for
non-negativity of $\Delta_1(\zeta)$ and $\Delta_2(\zeta)$. %
The positivity of $\Delta_1(\zeta)$ can be easily confirmed since
$\Delta_1(\zeta) = {\bm z}_1^T M_1 {\bm z}_1$ with $M_1 =
\mathrm{diag}(0.184, 7) \succ O$ 
and ${\bm z}_1 = [1, \zeta]^T$. 
On the other hand, $\Delta_2(\zeta)$ is expressed as $\Delta_2(\zeta) = {\bm z}_2^T
M_2 {\bm z}_2$ with monomials ${\bm
z}_2:=[1,\zeta,\zeta^2,\zeta^3,\zeta^4]^T$ 
and $M_2 = \mathrm{diag}(0.000555, 0.0315, \allowbreak0.192, -0.0382, 294)$. %
Thus, we examine the existence of a matrix $N_2$ such that $M_2 + N_2
\succ O$ and $\sum_{(j,k) \in \Theta_\ell} \nu_{jk}^{(2)} =
0~(\ell=2,3,\cdots,10)$ by solving the 
feasibility problem of the optimization program.
The optimization solver yields 
  \begin{align}
   N_2 =
   \begin{bmatrix}
    0 &  0 & -0.0264 &  0 & 0.0710 \\
    0 & 0.0528 & 0 & -1.3655 &  0.0012 \\
  -0.0264 & 0 &  2.5889 & -0.0012 & -16.797 \\
   0   & -1.3655 & -0.0012 & 33.595 & 0 \\
   0.0710 & 0.0012 & -16.797 & 0 & 0
   \end{bmatrix} 
  \end{align}
which indeed satisfies $M_2 + N_2 \succeq O$ and $\sum_{(j,k) \in \Theta_\ell}
 \nu_{jk}^{(2)} = 0~(\ell=2,3,\cdots,10)$.
 This implies that $\Delta_2(\zeta) = {\bm z}_2^T (M_2 + N_2)  {\bm z}_2 = {\bm z}_2^T M_2 {\bm z} \ge 0$ for all $\zeta \in \mathbb{R}$.
 Thus, the equilibrium point is locally stable, and the concentrations
 $[C_1, C_2]^T$ converge to the equilibrium when they are perturbed in the
 vicinity of $[C_1^*, C_2^*]$ (Fig. \ref{sim-fig}(B)).

 To see an example of an unstable case, we next consider the parameter sets used in
 Fig. \ref{sim-fig}(C), that is, $(b, v_1, v_2) = (0.055, 0, 0)$ (see Method section for the other parameters). 
 In this case, %
 \begin{align}
&  \Delta_1(\zeta) = 7\zeta^2 + 0.0679, \\
&  \Delta_2(\zeta) = 294\zeta^8 - 19.141 \zeta^6  - 0.0999 \zeta^4  + 0.0045
  \zeta^2 +0.000033.
 \end{align}
Thus, the sign of $\Delta_2(\zeta)$ determines the stability of the
equilibrium point. 
The SDP solver returns ``infeasible'', implying that there is no $N_2$ satisfying the conditions (ii) in Proposition 1.
This means that there exists a spatial frequency $\zeta$ for which the
characteristic polynomial  $\varphi(\zeta, s) = 0$ has a root in the
right-half complex plane.
Thus, the system (\ref{freq-eq}) is unstable for the frequency $\zeta$.
In particular, the destabilization occurs for some non-zero spatial frequency $\zeta$
 since $\Delta_1(\zeta) > 0$ and $\Delta_2(\zeta) > 0$ when $\zeta = 0$.
 This observation is consistent with the simulation result in
 Fig. \ref{sim-fig}(C) in that the steady state solution converges to
 the spatial periodic oscillations with some non-negative frequency. 

 \par
 \smallskip
Using the SDP based stability test, we further investigate how the advective flow
 affects the formation of spatial patterns in the reaction system (\ref{gray-scott-model}).  
Specifically, we first characterize the parameter space that induces
 transport-driven instability with and without flow
 (Fig. \ref{pmap} (A) and (B), respectively).
 In Fig. \ref{pmap}(B), the flow rates of $P$ and $Q$ were set $(v_1,
 v_2) = (0.3162d, 0.3162) = (1.897, 0.3162)$, respectively.
The difference of the flow rate is due to the Stokes-Einstein relation, 
 where the difference of the molecular size results in the diffusion and
 advection rates. 
Figures \ref{pmap}(A) and (B) illustrate that the advective flow broadens the parameter
 region for instability, implying that the formation of periodic patterns becomes more
 robust when there is advective flow in the system. 

 \par
 \smallskip
To further analyze the relation between the velocity of the flow and
 the parameter region for transport-driven instability, 
we define the flow rate as $(v_1, v_2) = (d v, v)$ and
 vary $v$ between 0 and 0.3162.
It should be noted that $v$ is the flow rate of $Q$ and is proportional to the velocity
 of the constant flow in the reactor (see Fig. \ref{sim-fig}(A)).
Fig. \ref{pmap}(C) illustrates the parameter region for transport-driven
 instability for different values of $v$, where $v = 0$ means that the
 transportation of the molecules is exclusively due to passive diffusion. 
 The result suggests that  the transport-driven instability is induced by advective flow once
 the speed of the flow reaches a threshold. 
 In particular, the instability region increases almost linearly with the
 speed of the flow and becomes almost twice as large as the diffusion
 only case when $v = 0.40$.
 The spatio-temporal simulations of the reaction-diffusion-advection system show that
periodic pattern formation does occur by increasing the speed of
 advection (Fig. \ref{pmap}(D-F)). 
 Although the link between diffusion-driven instability and biological
 pattern formation is often criticized due to the lack of robustness,
 these results suggest that the advective transportation of molecules could 
compensate for the fragility of the diffusion-driven pattern formation.

 \section{Pattern formation with specified spatial profiles}
In the design and analysis of reaction-diffusion-advection systems,
 we are often interested in the shape of spatial patterns, which is
 roughly determined by the spatial frequency, or the characteristic
 wavelength, of the spatial oscillations.
This requires identification of the destabilizing frequency, that is,
 the frequency $\zeta$ that makes the polynomials $\Delta_i(\zeta)$ negative.
In what follows, we extend Proposition 1 to enable local stability
analysis of reaction-diffusion-advection systems for a specified range
of spatial frequency $\mathcal{I} \subset \mathbb{R}$. %
Using the extended theorem, we can specify a range of spatial
frequency $\mathcal{I}$ that destabilizes the reaction-diffusion-advection
system. This facilitates the analysis and design of the spatial profiles
of transport-driven patterns.

\par
\smallskip
According to the Hurwitz condition presented in Section
\ref{LMI-sec}, the reaction-diffusion-advection system (\ref{linear-eq})
is stable for a given range of spatial frequency $\mathcal{I} \in
\mathbb{R}$ if and only if the polynomials 
$\Delta_i(\zeta)~(i=1,2,\cdots,n)$ is positive for all $\zeta \in
\mathcal{I}$.
To check the sign of $\Delta_i(\zeta)$ for a given rage $\mathcal{I}$, we
again introduce a set of linear matrix inequalities that is amenable to
semidefinite programming.

\medskip
\noindent
{\bf Proposition 2.}
Let $\mathcal{I}$ denote an open interval on real numbers, {\it i.e.,} 
$\mathcal{I} \subset \mathbb{R}$.
The following (i)--(iii) are equivalent.
\begin{itemize}
 \item[(i)] $\Delta_i(\zeta) \ge 0$ for all $\zeta \in \mathcal{I}$ and $i=1,2,\cdots,n$.
 \item[(ii)] 
There exist polynomials $f(\zeta)$ and $g(\zeta)$ such that $f(\zeta) \ge 0$ and $g(\zeta) \ge 0$ for all $\zeta \in \mathbb{R}$, and
$\Delta_i(\zeta) = f(\zeta) + h(\zeta) g(\zeta)$, where
 \begin{align}
 h(\zeta) := 
 \begin{cases}
  -(\zeta - \underline{\zeta})(\zeta - \overline{\zeta}) & {\rm when~}\mathcal{I}=[\underline{\zeta}, \overline{\zeta}] \\
 \zeta - \underline{\zeta} & {\rm when~}\mathcal{I}:=[\underline{\zeta}, \infty)\\
 -\zeta + \overline{\zeta} & {\rm when~}\mathcal{I}:=(-\infty, \overline{\zeta}]
 \end{cases}.
 \end{align} 
 \item[(iii)] The following conditions are satisfied.
 \begin{itemize}
 \item[$\bullet$] In the case of a finite interval $\mathcal{I} :=
  [\underline{\zeta}, \overline{\zeta}]$, define
	      $\delta_\ell^{(i)}$ as the coefficients of the polynomial
	      \begin{align}
	       \Delta_i \left(
	       \frac{(\overline{\zeta}-\underline{\zeta})\zeta +
	       (\overline{\zeta} + \underline{\zeta})}{2} \right) = \sum_{\ell} \delta_\ell^{(i)} \zeta^\ell.~~(i=1,2,\cdots,n)
\label{finite-delta}
	      \end{align}
	      Then, there exists $K_i \in \mathbb{R}^{(\ell_i+1) \times
	      (\ell_i+1)}$ and $L_i \in \mathbb{R}^{\ell_i \times \ell_i}$
	      such that
	       \begin{align}
	&	K_i \succeq O,~~L_i \succeq O, \label{KL-psd1}\\
&	       \delta_\ell^{(i)} = \sum_{(j,k) \in \Theta_{\ell+2}}
	       (\kappa_{jk}^{(i)} + \lambda_{jk}^{(i)}) - \sum_{(j,k) \in
	       \Theta_{\ell}} \lambda_{jk}^{(i)},
\label{cond3-eq1}
	       \end{align}
	      where $\ell_i := \lceil \mathrm{deg}(\Delta_i(\zeta))/2 \rceil$, and $\kappa_{jk}^{(i)}$ and $\lambda_{jk}^{(i)}$ represent the
	      $(j,k)$-th entry of the matrices $K_i$ and $L_i$, respectively. $\Theta_{\ell}$ is defined in Proposition 1.
 \item[$\bullet$] In the case of a semi-infinite interval $\mathcal{I} :=
  [\underline{\zeta}, \infty)$, define
	      $\delta_\ell^{(i)}$ as the coefficients of the polynomial
	      \begin{align}
	       \Delta_i \left(\zeta + \underline{\zeta} \right) = \sum_{\ell} \delta_\ell^{(i)} \zeta^\ell.~~(i=1,2,\cdots,n),
	       \label{inf-delta}
	      \end{align}
or in the case of $\mathcal{I} :=  (-\infty, \overline{\zeta}]$,
define $\delta_\ell^{(i)}$ as %
	      \begin{align}
	       \Delta_i \left(-\zeta + \overline{\zeta} \right) = \sum_{\ell} \delta_\ell^{(i)} \zeta^\ell.~~(i=1,2,\cdots,n).
	       \label{inf-delta2}
	      \end{align}
	      Then, there exists $K_i \in \mathbb{R}^{(\ell_i+1) \times
	      (\ell_i+1)}$ and a matrix $L_i$ such that
	       \begin{align}
		&	K_i \succeq O,~~L_i \succeq O, \label{KL-psd2}\\
		&	       \delta_\ell^{(i)} = \sum_{(j,k) \in \Theta_{\ell+2}}
	       \kappa_{jk}^{(i)} + \sum_{(j,k) \in
		\Theta_{\ell+1}} \lambda_{jk}^{(i)},
\label{cond3-eq2}
	       \end{align}
where $\ell_i := \lceil  \mathrm{deg}(\Delta_i(\zeta))/2 \rceil$, and
	      $\kappa_{jk}^{(i)}$ and $\lambda_{jk}^{(i)}$
	      represent the $(j,k)$-th entry of the matrices $K_i$ and
	      $L_i$, respectively.
	      The size of $L_i$ is $\ell_i+1$ by $\ell_i+1$ when
	      $\mathrm{deg}(\Delta_i(\zeta))$ is odd, and $\ell_i$ by
	      $\ell_i$ when $\mathrm{deg}(\Delta_i(\zeta))$ is even. $\Theta_{\ell}$ is defined in Proposition 1.
 \end{itemize}
\end{itemize}

It should be noted that $\Theta_\ell$ is an empty set for $\ell = 0, 1$. 
Similar to Proposition 1, the basic idea is to explore the SOS decomposition of $\Delta_i(\zeta)$ to guarantee its non-negativity 
on the interval $\mathcal{I}$ (see Supplementary Information for the proof). 
This problem essentially boils down to finding $f(\zeta)$ and $g(\zeta)$ in the condition (ii). %
In the condition (iii) of Proposition 2, we convert the condition (ii) into the form of matrix inequalities by using the change of variable. That is, the interval $\mathcal{I}$ is converted into $[-1,1]$ and $[0,\infty)$. This leads to the simple LMI based formulation (\ref{KL-psd1}), (\ref{cond3-eq1}), (\ref{KL-psd2}) and (\ref{cond3-eq2}), where the existence of the matrices $K_i$ and $L_i$ is equivalent to the existence of non-negative polynomials $f(\zeta) := {\bm z}_i^T K_i {\bm z}$ and $g(\zeta) := {\bm z}_i^T L_i {\bm z}$
such that $\Delta_i(\zeta) = f(\zeta) + h(\zeta) g(\zeta)$ with $h(\zeta) = -(\zeta-1)(\zeta+1)$ for the finite interval and $h(\zeta) = \zeta$ for the infinite interval case.
Since the existence problem of the matrices $K_i$ and $L_i$ can be implemented as a feasibility problem of a semidefinite program, it is possible to efficiently explore the matrices satisfying the constraints using existing solvers.

\par
\smallskip
The condition (ii) of Proposition 2 enables identification of the stable and
unstable spatial frequency of the reaction-diffusion-advection
system. 
This is particularly helpful to find the parameter space of the
reaction that leads to the formation of prespecified spatial patterns as demonstrated in the next example.

\medskip
\noindent
{\bf Remark 3.~}
When multi-dimensional space $\Omega$ is considered, the variable $\zeta$ in Proposition 2 is equal to the sum of the spatial frequency $\bar{\zeta} = \sum_{i=1}^{m} \zeta_i$ as shown in Remark 2. 
Thus, there can be multiple different combinations of frequency $\zeta_i~(i=1,2,\cdots,m)$ for a given destabilizing $\zeta$. 
This means that the growth rate of frequency components $\tilde{{\bm c}}({\bm \zeta}, t)$ are equal between all combinations of $\zeta_i$ that sums to $\zeta$. 
Therefore, any of these frequency components could appear as the resulting spatial pattern depending on the initial perturbation to the homogeneous steady state $\bar{\bm C}$.

 \begin{figure}[tb]
  \centering
\includegraphics[clip,width=12cm]{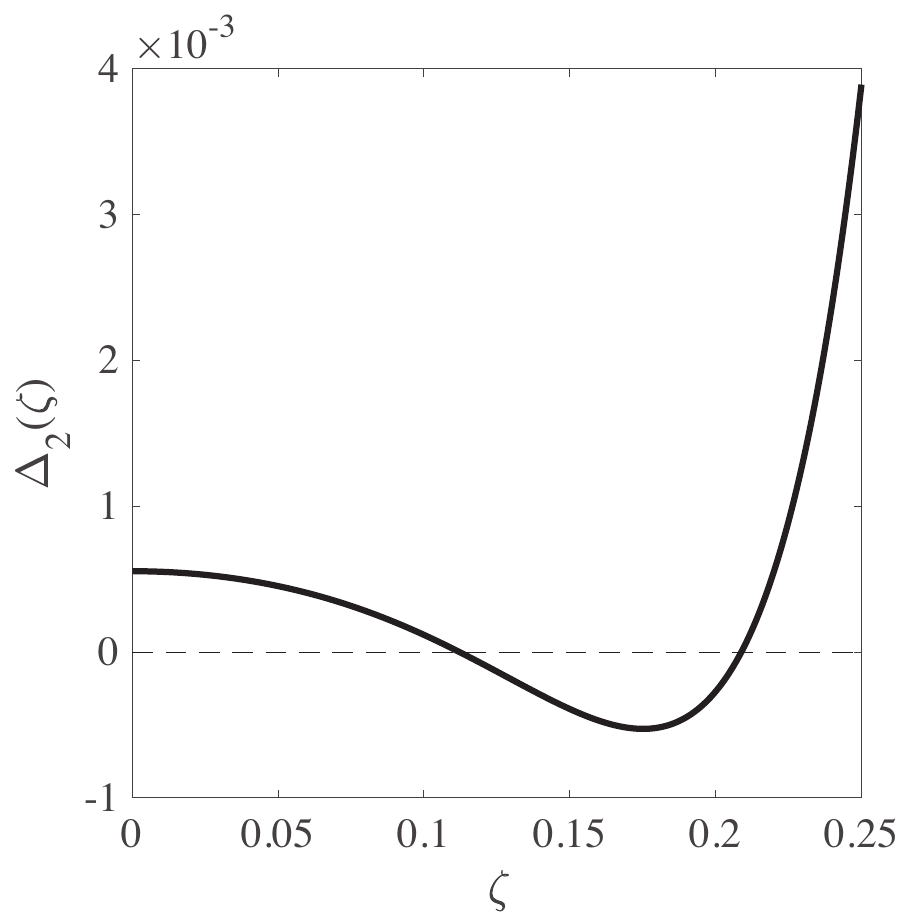}%
  \caption{{\bf Value of $\Delta_2(\zeta)$}. The
  reaction-diffusion-advection system is destabilized at the spatial
  frequency $\zeta$ satisfying  $\Delta_2(\zeta) < 0$.}
  \label{delta2-fig}
 \end{figure}

 \medskip
 \noindent
{\bf Example.}
We consider the reaction-diffusion-advection system (\ref{gray-scott-model}) and
its linearized model (\ref{linear-rd-eq}).
Let the parameters $a, b$ and $d$ be set as $(a,b,d) = (0.06, 0.04,
6.0)$ and $v = 0.3162$, with which the equilibrium is
 unstable as illustrated in  Fig. \ref{pmap}C (marked as ``F'').
As have already seen in the previous section, the matrix inequalities in
Proposition 1(ii)
are infeasible since $\Delta_2(\zeta)$ in (\ref{delta2-eq}) is negative
for some $\zeta \in \mathbb{R}$. This implies that the equilibrium
point of the reaction-diffusion-advection system is locally unstable, leading to
the formation of the spatial pattern in Fig. \ref{pmap}F.
To further analyze, we investigate the value of $\Delta_2(\zeta)$ as
shown in Fig. \ref{delta2-fig}. The figure implies that the reaction-diffusion-advection system is
destabilized around $0.12 \le \zeta \le 0.2$, whose 
wavelength corresponds to $31.4 \le 2\pi/\zeta \le 48.3$. 
We observe that the wavelength of the periodic spatial pattern in
Fig. \ref{pmap}F agrees with the destabilizing frequency. 

\par
\smallskip
Proposition 2 enables optimization-based analysis of the stabilizing/destabilizing frequency $\zeta$ without actually
computing $\Delta_i(\zeta)$ unlike Fig. \ref{delta2-fig}. %
As an example, we solve SDP with the condition (ii) in Proposition 2 by 
setting $\mathcal{I}=[0,0.1]$. 
Since $\Delta_2(\zeta)$ is positive on $\mathcal{I}$, the SDP is
feasible and the solver finds 
\begin{eqnarray*}
 & K_2 = 10^{-4} \times
  \begin{bmatrix}
     2.6498  &  0.5060 &  -11.327   & -0.7936  &  9.7421\\
    0.5060 &   9.7828  & -8.9253 & -12.081 &  10.089\\
  -11.327 &  -8.9253 &  79.539 &  11.580 & -76.375\\
   -0.7936 & -12.081 &  11.580 &  15.730 &  -13.547\\
   9.7421  & 10.089 & -76.375 & -13.547 &  75.743
  \end{bmatrix}
 \succeq O,\\
 & L_2 = 10^{-4} \times
   \begin{bmatrix}
   1.8741  & -1.5595 &  -4.2902  &  3.3575\\
   -1.5595 &  22.153 &   4.7537 & -38.461\\
   -4.2902 &   4.7537 &  15.638  & -13.547\\
    3.3575 & -38.461 & -13.547 &  75.743
   \end{bmatrix}
 \succeq O,\\ 
\end{eqnarray*}
which verifies non-negativity of $\Delta_2(\zeta)$ for $0 \le \zeta \le
0.1$.
Similarly, for $\mathcal{I}=[0.25, \infty)$, the solver finds 
\begin{eqnarray*}
 & K_2 =
  \begin{bmatrix}
    0.0039 &  -0.1259  &  0.1373 &  0.1207 &  0.0830\\
   -0.1259 &   8.2473 & -22.293 & -12.915 & -10.193\\
    0.1373 & -22.293 & 139.67 &  52.697 &  75.458\\
    0.1207 & -12.915 &  52.697 & 264.83 & 166.16\\
    0.0830 & -10.193 &  75.458 & 166.16 & 294.00
  \end{bmatrix}
  \succeq O,\\
 & L_2 = 
\begin{bmatrix}
 0.4259 &  -2.9496 &  1.0891  &  0.9705\\
   -2.9496 & 62.322 & -11.063 &  12.481\\
    1.0891 & -11.063 & 169.72 &  56.854\\
 0.9705 &  12.481 &  56.854 & 255.68
\end{bmatrix}
\succeq O.
\end{eqnarray*}
On the other hand, the solver certifies that there is no $K_i$ and $L_i$
that satisfies conditions (ii) in Proposition 2 when $0.1 < \zeta < 0.25$.
These results imply that $\Delta_2(\zeta)$ is positive for $\zeta \in [0,0.1]$
and $\zeta \in [0.25, \infty)$ and is negative for some $\zeta \in (0.1,
0.25)$. Thus, the destabilizing frequency is identified as $0.1 <
\zeta < 0.25$ %
without explicitly computing the value of $\Delta_2(\zeta)$.
The SDP feasibility test introduced above allows the identification of
the interval of destabilizing frequency.
Using this feature, it is possible to further narrow down the parameter sets for
diffusion/advection-driven instability (Fig. \ref{pmap}(A-C)) with 
the additional constraints of the wavelength of spatial patterns.

\par
\smallskip
When there is a single connected interval of destabilizing frequency $\mathcal{I}$, we can also use a bisection-type algorithm to 
precisely estimate the destabilizing frequency. 
Specifically, we first solve the SOS program in Proposition 2 for some initial interval $[0, \underline{\zeta}]$ with sufficiently large $\underline{\zeta}$, and iteratively divide $\underline{\zeta}$ into a half until the SOS program in Proposition 2 returns a feasible solution. Once we find a frequency that gives a feasible solution, say $\zeta_*$, the lower bound of the destabilizing frequency is between $\zeta_*$ and $2 \zeta_*$. Thus, we further solve the SOS program by setting $\underline{\zeta} = 1.5 \zeta_*$ and divide the domain into a half until it becomes small enough, which converges to the lower bound of the destabilizing frequency.
The upper bound of the destabilizing frequency can be found by essentially the same idea, where we initially solve the SOS program in Proposition 2 for $[\overline{\zeta}, \infty)$ with sufficiently small $\overline{\zeta}$, and iteratively narrow the domain by $[2\overline{\zeta}, \infty)$. 
It should, however, be noted that this bisection-type algorithm gives only the upper and lower bounds of destabilizing frequency when there are multiple destabilizing frequency intervals. For example, $\zeta_1$ and $\zeta_4$ are obtained when there are two disconnected intervals of destabilizing frequency $\mathcal{I}_1 = (\zeta_1, \zeta_2)$ and $\mathcal{I}_2 = (\zeta_3, \zeta_4)$ with $\zeta_2 < \zeta_3$.

\section{Conclusion and Discussion}
Self-organizing phenomena in spatially interacting chemical systems have
been studied in biology and chemistry for a long time due to its link with biological developmental process. 
In microbiology, it is known that culture conditions may affect the spatial profile of bacterial colonies, and its link with reaction-diffusion systems have been extensively studied (see Chapter 11 of \cite{Murray2003} for example). 
In developmental biology, experimental works revealed that spatio-temporal molecular patterning in embryos affects morphology, the structure of organisms \cite{Kondo2010,Palmeirim1997,Bessho2001}. This motivates spatio-temporal control of morphogen concentrations in tissue engineering \cite{Biondi2008}. 
In addition to the biological applications, there are potential engineering applications of molecular based communications, where 
spatially dispersed micro and nano-scale systems communicate with each other by small diffusible molecules \cite{Nakano2014}.
Therefore, understanding the dynamics of transport-driven pattern formation will not only contribute to biological science but also helps open up new engineering applications.

\par
\smallskip
This paper has proposed a novel computational approach to 
analyzing the local stability/instability of reaction-diffusion-advection systems.
The proposed algebraic conditions in Propositions 1 and 2 bypass the
iterative stability analysis of individual Fourier components and
provide a route toward the direct characterization of the local behavior by a single run of mathematical
optimization, speeding the analysis and synthesis of
self-organizing chemical systems in biology and chemistry.
The condition in Proposition 2 further allows the computation of 
 the destabilizing spatial frequency. This helps detailed quantitative analysis of the
 spatio-temporal profile of the self-organizing chemical
 concentrations.
We have numerically illustrated the optimization-based 
analysis process using the auto-catalytic reaction model with diffusion
and advection.

 \par
 \smallskip
From a theoretical viewpoint, our development hinges upon the sum-of-squares (SOS) optimization technique
 \cite{Parrilo2003} to prove non-negativity of $\Delta_i(\zeta)$, 
 which implies stability of the reaction-diffusion-advection systems.
In the last decade, the SOS optimization was extensively studied in
control engineering community to analyze the stability of nonlinear 
dynamical systems. Software packages were developed to facilitate the implementation of SOS
 optimization programs \cite{SOSTOOLS}. 
In general, the existence of SOS decomposition of a polynomial implies 
 non-negativity, but the converse is not necessarily the case. In other
 words, not all non-negative polynomials can be represented as a sum of
 squares (see \cite{Parrilo2003} for example).
In our theoretical development, however, we have used the fact that 
the polynomial inequalities associated with the stability analysis are univariate, 
 in which case there always exists an SOS  decomposition if the polynomial
 is non-negative, leading to not only the sufficient but also the necessary condition for the stability
 of reaction-diffusion-advection systems. This is particularly
 useful for the study of transport-driven pattern formation since the
 condition can provide rigorous instability certificate of the reaction-diffusion-advection systems. 
It is also worth mentioning that the polynomial inequality conditions $\Delta_i(\zeta) \ge 0~(i=1,2,\cdots,n)$ are univariate even 
for multi-dimensional reaction-diffusion-advection systems as discussed in Remark 2. 
This is attractive from a viewpoint of computational burden since, in general, the size of the matrices involved in the SOS programs such as $M_i, N_i, K_i$ and $L_i$ grow in a combinatorial fashion with the number of variables and the degrees of polynomials, which is one of actively studied issues in recent years \cite{Waki2006, Ahmadi2017, Weisser2018}. 
A limitation of the proposed approach is that it can only verify local properties around a given equilibrium point due to the linearization, and thus, it cannot 
rigorously guarantee the emergence of inhomogeneous spatial patterns in nonlinear reaction-diffusion-advection systems. 
To overcome this issue, an alternative approach would be to use Lyapunov's approach to obtain sufficient conditions for global stability \cite{Jovanovic2008,Arcak2011,Shafi2013,Valmorbida2016}.

\par
\smallskip
A major criticism about Turing's diffusion-driven instability
 is that the parameter space for the instability is too small to achieve even in engineered chemical systems
although spatial pattern formation is quite a common phenomenon in biology. 
Recent theoretical works, on the other hand, presented that stochastic intrinsic noise of
chemical reactions in biological cells enhances diffusion-driven
instability, resulting in the increase of the instability parameter
regime \cite{Biancalani2010,Butler2011,Scott2011,HoriCDC2012}. 
Our transport-driven instability analysis in Fig. \ref{pmap}(A)-(C) suggests 
that advective flow is another factor that broadens the parameter space
for transport-driven instability. In fact, Fig. \ref{pmap}F shows that the originally stable
reaction-diffusion system (Fig. \ref{pmap}D) exhibits spatially
inhomogeneous patterns by adding the flow.
These results imply that the robustness of self-organization in nature
could be guaranteed by circulatory systems in
addition to intrinsic noise.

\par
\smallskip
The idea of flow-driven instability was previously presented in 
\cite{Rovinsky1992,Malchow2000,Gholami2015,Vidal-Henriquez2017}, where periodic spatial
patterns were observed due to the
destabilizing effect of advection. 
The theoretical prediction was also verified with engineered chemical
and biological systems \cite{Rovinsky1993,Eckstein2018}. 
Unlike the model (\ref{gray-scott-model}) in this paper, the
previous analyses were limited for systems where only one molecule could
diffuse and flow while other molecules were immobilized.
This assumption can be easily relaxed with the proposed stability analysis
method as it is capable of dealing with arbitrarily many diffusing and
flowing molecules in principle despite the complex eigenvalues of the advection
operator.
Although this paper has only shown a simple reaction example to focus on
 the verification and demonstration of the theoretical development,
it will be helpful to examine more biologically relevant models, in
 future, to understand the underlying mechanisms of the flow-driven robustification of the spatial pattern formation. 
The authors also envision that there are potential extensions of the proposed SOS programs to robust stability analysis and (reaction) design problems, given that these topics are actively studied using SOS programs in control systems community \cite{Papachristodoulou2005}.

\section*{Method}
The spatio-temporal dynamics in Fig. \ref{sim-fig}(B)-(D) and
Fig. \ref{pmap}(D)-(F) were simulated with the periodic boundary
condition using Wolfram Mathematica 11.0.1.0 on macOS 10.12.6.
These colormaps illustrate the concentration of $P$, or $C_1(x, t)$.
In all simulations, $a = 0.06$ and $d = 6.0$ were used. The spatial
length was $L = 30\pi$.
Figure \ref{pmap}(D) and (F) are repeated from Fig. \ref{sim-fig}(B) and
(D), respectively, for the clarity of presentation.
For Fig. \ref{pmap}(E), the parameters were set $a=0.06, b=0.04, d=6.0, v_1=1.518, v_2=0.2530$.
For all spatio-temporal simulations, the initial values were set 
$C_1(x,0) = 0.5 + 0.0025 \sum_{k=1}^{20} \{\cos(2k\pi x/L) + \sin(2k\pi x/L)\}$
 and $C_2(x,0) = 0.2 + 0.0025\sum_{k=1}^{20} \{\cos(2k\pi x/L) + \sin(2k\pi x/L)\}$.

\par
\smallskip
The semidefinite programs were run with SeDuMi 1.3 \cite{Strum1999} and
YALMIP toolbox \cite{Lofberg2004} on MATLAB 2016b.

 \section*{Data, codes and materials}
The program codes used in this study are available at GitHub
 (\url{https://github.com/hori-group/SDP_for_flow-driven_stability_analysis})

 \section*{Competing interests}
 The authors have no competing interests.

 \section*{Authors' contributions}
 Y.H. conceived of and designed the study.
 Both authors performed the mathematical derivation and implemented the
 computational codes for optimization and simulations. 
Both authors drafted the manuscript and gave final approval for publication.

 \section*{Funding}
This work was supported in part by JSPS KAKENHI Grant Number JP18H01464
and 15J09841.

\bibliographystyle{vancouver}

\end{document}